\documentclass[aps,prx,twocolumn,groupedaddress]{revtex4-2}

\usepackage{amsmath} 
\usepackage{graphicx}
\usepackage{xcolor}
\usepackage{multirow}
\usepackage{adjustbox}
\usepackage{comment}
\usepackage{multirow}
\usepackage[T1]{fontenc} 
\usepackage{amssymb}

\usepackage{tikz}
\usetikzlibrary{arrows.meta}
\newcommand{\longdashleftrightarrow}[1][1.75em]{\mathrel{
	\tikz[baseline]{\draw[dash pattern=on .25em off .1 em,<->](0,.58ex)--(#1,.58ex)}}}
\newcommand{\longdashleftarrow}[1][1.75em]{\mathrel{
		\tikz[baseline]{\draw[dash pattern=on .25em off .1 em,<-](0,.58ex)--(#1,.58ex)}}}

\newcommand*{\blue}[1]{\textcolor{blue}{\ensuremath{\bar{#1}}}}
\newcommand*{\red}[1]{\textcolor{red}{\ensuremath{#1'}}}
\newcommand*{\green}[1]{\textcolor{green}{\ensuremath{\bar{#1}'}}}

\begin{document}

\title{Algebra of symmetry allowed physical properties of crystals}

\author{Piotr Fabrykiewicz}
\email[]{p.fabrykiewicz@fz-juelich.de}

\affiliation{J{\"u}lich Centre for Neutron Science at Heinz Maier-Leibnitz Zentrum, Forschungszentrum Jülich GmbH, Lichtenbergstr. 1, 85747 Garching, Germany}
\affiliation{Institute of Crystallography, RWTH Aachen University, J{\"a}gerstr. 17-19, 52066 Aachen, Germany}

\date{\today}

\begin{abstract}
	The symmetry of a physical system given by the 122 magnetic point groups (MPGs) leads to only 21 sets of restrictions on any physical tensors. Those, so called, tenor forms were labeled by letters from A to U by R. R. Birss (Symmetry and Magnetism, North Holland, Amsterdam, 1964). A new concept of property-related point groups (PRPGs) is proposed. It is shown that the 21 PRPGs isomorphic to the 21 non-centrosymmetric crystallographic point groups (CPGs) corresponds to the 21 tensor forms. Therefore instead of a symbols from A to U one can use PRPGs as mathematical objects corresponding to the tenor forms. From the Neumann's principle given multipole component is only allowed in a system when a group describing symmetry of the system is a subgroup of the multipole component symmetry. Each tensor can be expressed in the terms of the multipole degrees of freedom, i.e. it corresponds to a subset of all multipole components allowed by given PRPG. Those relations leads to the algebra of tensor forms, where space of allowed tensor components and space of PRPG operators are somehow dual to each other.
\end{abstract}

\maketitle


\section{Introduction}

\begin{figure}
	\includegraphics[width=0.40\hsize]{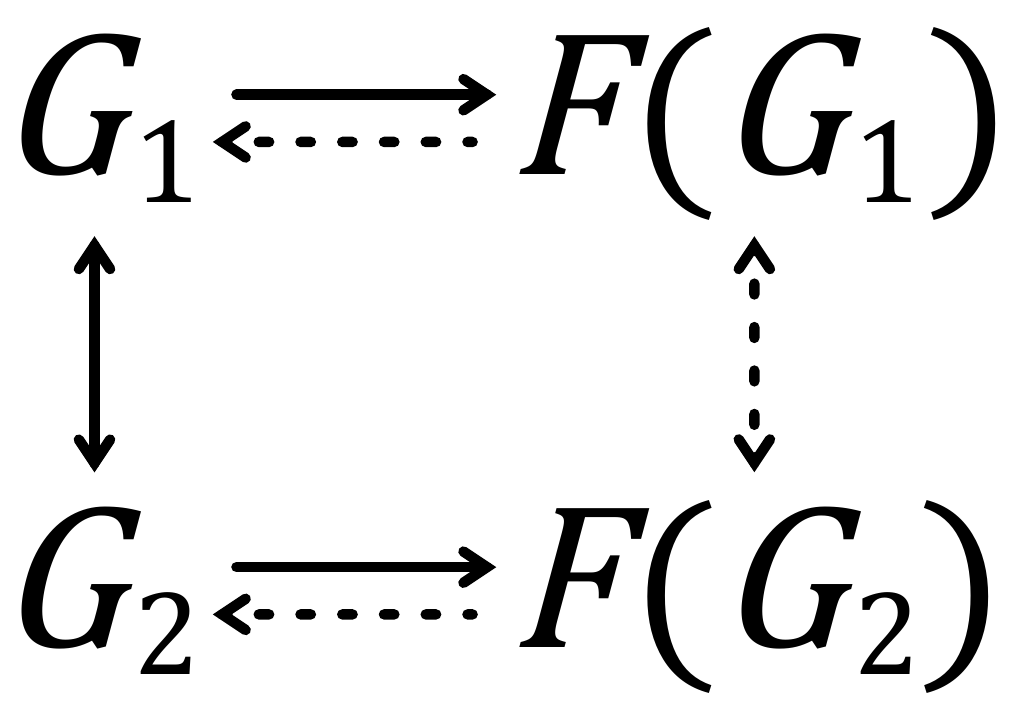}
	\caption{\label{Fig_graph}
		Relation between groups $G_1$, $G_2$ and the symmetry allowed forms of physical properties $F(G_1)$, $F(G_2)$. The solid arrows $\longrightarrow$ mean the application of the Neumann's principle~\cite{Neumann_book, Voigt_book, Minnigerode_1884, Curie1884, Shubnikov_Koptsik_book} to multipole~\cite{Hayami2018, Yatsushiro2021} or tensor~\cite{Nye_book, Birss1963, Birss_book, Sirotin_Shaskolskaya_book, MTENSOR} classifications. The dashed arrows $\longdashleftarrow$ denote the reverse to these classifications. The symbol $\updownarrow$ represents relations between groups. The solid arrows represent fully solved procedures, while the dashed arrows indicate a partially solved problem which can be further developed.
	}
\end{figure}

In a study of the properties of multiferroic materials as well as other emerging phenomena in condensed matter physics it is important to establish the property-symmetry relationship~\cite{Neumann_book, Voigt_book, Minnigerode_1884, Curie1884, Shubnikov_Koptsik_book}. An important step is to use the restrictions of the symmetry groups on tensors and multipoles. In early classifications only the 32 crystallographic point groups (CPGs) were considered, e.g.~\cite{Nye_book, Hayami2018}. In next decades, the classifications were extended to the 122 magnetic point groups (MPGs)~\cite{Birss_book, Yatsushiro2021}. The introduction of the time-reversal operation $1'$ in MPG makes the tensors and multipoles classification longer and more complicated than for 32 CPGs. There are many classifications of symmetry restrictions on tensors~\cite{Nye_book, Landau_Lifshitz_book, Sirotin_Shaskolskaya_book, Birss1963, Birss_book, Shtrikman1965, Kleiner1966,  Cracknell1973, Kopsky1976, Litvin1991, Eremenko_book, Grimmer1993, Grimmer1994, Grimmer2017, Kopsky2015, MTENSOR} and multipoles~\cite{Hayami2018, Yatsushiro2021}. Those classifications give answer to the question: {\em What parameters of the physical property $F(G)$ are allowed by the group $G$?} This is shown in a symbolic way by $G\longrightarrow F(G)$ in Figure~\ref{Fig_graph}. The reverse question: {\em What groups allow exactly for this specific $F(G)$?}, shown with the symbol $G\longdashleftarrow F(G)$ is only partially solved in available studies~\cite{Hayami2018, Yatsushiro2021, Nye_book, Birss1963, Birss_book, Sirotin_Shaskolskaya_book, MTENSOR} and this question can be further developed. Can the relation between groups $\updownarrow$ be extrapolated to the relation between symmetry allowed properties $\raisebox{0.3ex}{\rotatebox[origin=c]{-90}{\scalebox{0.55}{$\longdashleftrightarrow$}}}$~? This is the main question of the present paper.

\begin{table*}
	\caption{\label{Table:types} Definitions of types of physical properties with their behavior under antisymmetries \blue{1}, \red{1} and \green{1}. "$+$" stands for the invariance. "$-$" stands for sign reversal. $l$ and $N$ stands for the rank of multipoles and tensors, respectively. 
		In the present paper the notation from Sirotin \& Shaskolskaya~\cite{Sirotin_Shaskolskaya_book} (see column 1) is used. Other notations from e.g. Hayami \textit{et al.} \cite{Hayami2018}, Winkler \& Z\"ulicke~\cite{Winkler2025}, Birss~\cite{Birss_book}, Grimmer~\cite{Grimmer1991},  Kopsk\'{y}~\cite{Kopsky1976} are given for reference.}
	\begin{ruledtabular}
		\begin{tabular}{rllllcccccc}
			                                               &     &          &         &            &                 &                                             &                                       \multicolumn{4}{c}{\textrm{Tensors}}                                       \\ \cline{8-11}
			                       \multicolumn{5}{c}{\textrm{Definition}}                         & \multicolumn{2}{c}{Multipoles~\cite{Hayami2018, Winkler2025}} & \multicolumn{2}{c}{\textrm{Birss~\cite{Birss_book}}} & Grimmer~\cite{Grimmer1991} & Kopsk\'{y}~\cite{Kopsky1976} \\ \cline{1-5}\cline{6-7}\cline{8-9}
			\textrm{Type~\cite{Sirotin_Shaskolskaya_book}} & 1   & \blue{1} & \red{1} & \green{1} & \text{even }$l$ &               \text{odd }$l$                & \textrm{even N} &           \textrm{odd N}           &           any N            &            any N             \\ \hline
			                                   \text{Even} & $+$ & $+$      & $+$     & $+$        &        E        &                     ET                      & polar i-tensor  &           axial i-tensor           &          i-tensor          &        $1^+_e$-tensor        \\
			                               \text{Magnetic} & $+$ & $+$      & $-$     & $-$        &       MT        &                      M                      & polar c-tensor  &           axial c-tensor           &          s-tensor          &        $1^+_m$-tensor        \\
			                               \text{Electric} & $+$ & $-$      & $+$     & $-$        &       ET        &                      E                      & axial i-tensor  &           polar i-tensor           &          t-tensor          &        $1^-_e$-tensor        \\
			                        \text{Magnetoelectric} & $+$ & $-$      & $-$     & $+$        &        M        &                     MT                      & axial c-tensor  &           polar c-tensor           &          u-tensor          &        $1^-_m$-tensor
		\end{tabular}
	\end{ruledtabular}
\end{table*}

Opechowski~\cite{Opechowski1974} based on work of Ascher~\cite{Ascher1966} noted that the number of MPGs allowing for ferromagnetic, ferroelectric, and ferrotoroidal moments are all equal 31. Opechowski stated that the same number of groups for the three different phenomena is not due to a coincidence, but it is a sign of the internal mathematical structure of the MPGs. Kopsk\'y~\cite{Kopsky1976, Kopsky2006} extended this observation to physical property tensors by analyzing the structure of MPGs.

In a similar approach it was shown that all kinds of long-range magnetic ordering in crystals, described by the 1651 magnetic space groups, can be reduced to only 64 magnetic modes~\cite{Fabrykiewicz2021}. In full analogy to ordered (ferro- or antiferro-) magnetic orderings, the (ferro- or \mbox{antiferro-)} electric orderings can be also reduced to only 64 modes, i.e. to 64 types of Wyckoff positions. The analogy can be extended to (ferro- or antiferro-) toroidal orderings which can be also described by using 64 modes as well~\cite{Fabrykiewicz2023a}.

The main line of the present paper is based on a similar quantitative observation. When the symmetry of a physical system is given by one of the 122 magnetic point groups (MPG) then the tensors which describe the symmetry allowed phenomena are limited to 21 tensor-forms enumerated by Birss~\cite{Birss_book} and labeled by letters from A to U.
There are 32 (non-magnetic) crystallopraphic point groups (CPG) and 21 out of them are non-centrosymmetric. It will turn-out that there is a one-to-one correspondence between the 21 non-centrosymmetric CPGs and the 21 tensor forms~\cite{Litvin1994}.

Following the approach from~\cite{Fabrykiewicz2023a}, each magnetic point group operation $\hat{O}$ has an unambiguous decomposition:
\begin{equation}
	\hat{O} = \hat{R} \cdot \hat{I} \label{Eq:RI}
\end{equation}
into a spatially anisotropic component (proper rotations):
\begin{equation}
	\hat{R} \in \{ 1, 2, 3, 4, 6 \} \hspace*{1em} \label{Eq:R}
\end{equation}
and spatially isotropic component (antisymmetries):
\begin{equation}
	\hat{I} \in \{ 1, \blue{1}, \red{1}, \green{1} \}, \label{Eq:I}
\end{equation}
where space-, time- and combined inversions are defined as:
\begin{subequations}
	\begin{align}
		\blue{1}\hspace*{0.3em} (x, y, z, t) &= (-x, -y, -z, \textcolor{white}{+}t),\\
		\red{1} (x, y, z, t) &= (\textcolor{white}{+}x, \textcolor{white}{+}y, \textcolor{white}{+}z, -t),\\
		\green{1} (x, y, z, t) &= (-x, -y, -z, -t).
	\end{align}
\end{subequations}

The action of $\hat{R}$ and $\hat{I}$ on a physical property can be classified separately, because those components commute ($\hat{R} \cdot \hat{I} = \hat{I} \cdot \hat{R}$). Each antisymmetry has a distinct physical meaning, however, mathematically they behave the same way in the structure of magnetic point groups~\cite{Kopsky1976}. As the square of any antisymmetry is equal to identity, a physical property can be only invariant ($+$) or change ($-$) sign upon an action of antisymmetry as given in Table~\ref{Table:types}.  In the literature one can find two conventions of naming the multipole and tensor types. In one convention used e.g. by Birss~\cite{Birss_book}, Hayami \textit{et al.}~\cite{Hayami2018} and Winkler \& Z\"ulicke~\cite{Winkler2025} the symbol of the multipole/tensor type depends explicitly on the parity of the rank ($l$ or $N$, see Table~\ref{Table:types}). In a second convention used by e.g. Kopsk\'{y}~\cite{Kopsky1976}, Sirotin \& Shaskolskaya~\cite{Sirotin_Shaskolskaya_book} and Grimmer~\cite{Grimmer1991} the symbol is the same for all ranks. Moreover, the above mentioned papers~\cite{Birss_book, Kopsky1976, Sirotin_Shaskolskaya_book, Grimmer1991, Hayami2018, Winkler2025} use different names of the physical property types. The present paper will follow the notation from Sirotin \& Shaskolskaya~\cite{Sirotin_Shaskolskaya_book} with four types of physical properties: even, magnetic, electric and magnetoelectric, see first column in Table~\ref{Table:types}.


\section{Property-related point groups}

\subsection{Mapping of MPGs into PRPGs}

Let us consider one group out of 122 MPGs and one out of the four types of physical properties. All antisymmetries to which this property is invariant are indistinguishable from identity, 1, and all antisymmetries which reverse the sign of the property are indistinguishable from each other. One can replace the former ones with identity, 1, and the latter ones with the abstract property-related antisymmetry, $1^\dagger$ in both symmetry of property and group describing symmetry of physical system. This procedure starting from MPG defines a new property-related point group (PRPG). In Fig.~\ref{Fig_graph} the operator $\longdashleftarrow$ can lead to a set with more than one MPG, while for PRPGs there is one to one correspondence between $G$ and $F(G)$.  The concept of replacing the selected MPG with a PRPG means in simple terms {\em to focus on what is important and forget what is unimportant}.

The idea of the property-related point groups (PRPGs) is based on the paper by Litvin~\cite{Litvin1994} with following abstract: {\em A method is presented to determine the form of magnetic physical-property tensors from the form of non-magnetic physical-property tensors.} Litvin~\cite{Litvin1994} related forms of all four types of tensors in MPGs with forms of even and electric tensors in CPGs. PRPGs are isomorphic to CPGs can be viewed as an intermediate step in the Litvin's procedure. The advantage of PRPGs is that they always describe single type of property, namely $1^\dagger$-odd properties.

MPG is mapped into PRPG in the procedure shown in Table~\ref{Table_types}. Any physical property is invariant to the subgroup of the group consisting of all antisymmetries. One can use this group as a kernel of the epimorphism of MPGs into PRPGs, by replacing $\hat{I}$ component in all operations $\hat{O} = \hat{R} \cdot \hat{I}$ of MPG by identity $1$ or the property related antisymmetry $1^\dagger$ depending on the property type according to lower part of Table~\ref{Table_types}.

PRPGs are collected in Table~\ref{Table_groups}. From the 32 PRPGs only 21 PRPGs are physically important as the remaining ones contain $1^\dagger$ as generator, and therefore turn each $1^\dagger$-odd property to zero. Realizations of the above procedure for each property type in each of the 122 magnetic point groups is given in Table S1 in the Supplemental Material (SM)~\cite{SM_joint}. As it will be shown below, the use of PRPG will lead to a shortening of the classifications because (i) instead of 122 MPG one needs only 21 PRPG and (ii) instead of considering four property types one needs only one $1^\dagger$-odd property type. Depending on the property-type, each PRPG is decoded into a set of in general different MPGs as shown in Table S2 in SM~\cite{SM_joint}, which can be viewed as a reversed Table S1. MPGs restrictions to the property before this morphism are the same as PRPGs restrictions to the $1^\dagger$-odd property with unchanged spatial symmetry. Mapping of MPGs into PRPG can be done easily when RI notation~\cite{Fabrykiewicz2023a} of MPG is used. The dictionary between international and RI notations for all magnetic point groups is given in Table S3 in SM~\cite{SM_joint}.

Replacing property related antisymmetry with the spacial inversion $1^\dagger \rightarrow \blue{1}$ creates an isomorphism between PRPGs and CPGs as shown in Table~\ref{Table_groups}, e.g. $m\rightarrow2^\dagger$. The physically important 21 PRPGs are isomorphic to 21 non-centrosymmetric CPGs. This allows to transfer symmetry restrictions of electric-type properties in crystallographic point groups to $1^\dagger$-odd properties in property-related point groups. Combining this isomorphism with epimorphism of MPGs into PRPGs allows us to read symmetry restrictions on magnetic and electromagnetic type properties from CPGs.

\begin{table}
	\begin{ruledtabular}
		\caption{\label{Table_types} Upper part: Types of properties. "$+$" stands for the invariance. "$-$" stands for the sign reversal. Lower part: Rules how to obtain a PRPG from an MPG.}
		\begin{tabular}{ccccc}
			&         1         &                \blue{1}                 &                \red{1}                 &                \green{1}                 \\ \hline
			even       &        $+$        &                   $+$                   &                  $+$                   &                   $+$                    \\
			magnetic     &        $+$        &                   $+$                   &                  $-$                   &                   $-$                    \\
			electric     &        $+$        &                   $-$                   &                  $+$                   &                   $-$                    \\
			electromagnetic &        $+$        &                   $-$                   &                  $-$                   &                   $+$                    \\ \hline
			even       & $1 \rightarrow 1$ & $\blue{1} \rightarrow 1$\hspace{0.45em} & $\red{1} \rightarrow 1$\hspace{0.45em} & $\green{1} \rightarrow 1$\hspace{0.45em} \\
			magnetic     & $1 \rightarrow 1$ & $\blue{1} \rightarrow 1$\hspace{0.45em} &    $\red{1} \rightarrow 1^\dagger$     &    $\green{1} \rightarrow 1^\dagger$     \\
			electric     & $1 \rightarrow 1$ &    $\blue{1} \rightarrow 1^\dagger$     & $\red{1} \rightarrow 1$\hspace{0.45em} &    $\green{1} \rightarrow 1^\dagger$     \\
			electromagnetic & $1 \rightarrow 1$ &    $\blue{1} \rightarrow 1^\dagger$     &    $\red{1} \rightarrow 1^\dagger$     & $\green{1} \rightarrow 1$\hspace{0.45em}
		\end{tabular}
	\end{ruledtabular}
\end{table}

\begin{table*}
	\begin{ruledtabular}
		\caption{\label{Table_groups} The 32 property-related point groups (PRPGs) (left columns) isomorphic to crystallographic point groups (CPGs) (right columns) are collected within crystallographic systems (headers). Rows with groups starts with PRPG type. Analogous table for 122 magnetic point groups (MPGs) is given in Supplemental Material (SM)~\cite{SM_joint} as Table S7.}
		\begin{tabular}{lllllllllll}
			Type                & PRPG                  & CPG         & PRPG                  & CPG         & PRPG                  & CPG       & PRPG          & CPG          & PRPG                  & CPG         \\ \cline{1-1}\cline{2-3}\cline{4-5}\cline{6-7}\cline{8-9}\cline{10-11}
			&                       &             &                       &             &   \multicolumn{2}{c}{Triclinic}   & \multicolumn{2}{c}{Trigonal} &      \multicolumn{2}{c}{Cubic}      \\ \cline{6-7}\cline{8-9}\cline{10-11}
			White               &                       &             &                       &             & $1$                   & $1$       & $3$           & $3$          & $23$                  & $23$        \\
			Gray                &                       &             &                       &             & $1^\dagger$           & $\bar{1}$ & $31^\dagger$  & $\bar{3}$    & $231^\dagger$         & $m\bar{3}$  \\ \cline{2-3}\cline{4-5}\cline{6-7}
			&    \multicolumn{2}{c}{Hexagonal}    &   \multicolumn{2}{c}{Tetragonal}    &  \multicolumn{2}{c}{Monoclinic}   &               &              &                       &             \\ \cline{2-3}\cline{4-5}\cline{6-7}
			White               & $6$                   & $6$         & $4$                   & $4$         & $2$                   & $2$       & $32$          & $32$         & $432$                 & $432$       \\
			B\&W$^\text{iso}$   & $6^\dagger$           & $\bar{6}$   & $4^\dagger$           & $\bar{4}$   & $2^\dagger$           & $m$       & $32^\dagger$  & $3m$         & $4^\dagger32^\dagger$ & $\bar{4}3m$ \\
			Gray                & $61^\dagger$          & $6/m$       & $41^\dagger$          & $4/m$       & $21^\dagger$          & $2/m$     & $321^\dagger$ & $\bar{3}m$   & $4321^\dagger$        & $m\bar{3}m$ \\ \cline{6-7}
			&                       &             &                       &             & \multicolumn{2}{c}{Orthorhombic}  &               &              &                       &             \\ \cline{6-7}
			White               & $622$                 & $622$       & $422$                 & $422$       & $222$                 & $222$     &               &              &                       &             \\
			B\&W$^\text{iso}$   & $62^\dagger2^\dagger$ & $6mm$       & $42^\dagger2^\dagger$ & $4mm$       &                       &           &               &              &                       &             \\
			B\&W$^\text{aniso}$ & $6^\dagger22^\dagger$ & $\bar{6}2m$ & $4^\dagger22^\dagger$ & $\bar{4}2m$ & $2^\dagger2^\dagger2$ & $mm2$     &               &              &                       &             \\
			Gray                & $6221^\dagger$        & $6/mmm$     & $4221^\dagger$        & $4/mmm$     & $2221^\dagger$        & $mmm$     &               &              &                       &
		\end{tabular}
	\end{ruledtabular}
\end{table*}


\subsection{Structure of MPGs and PRPGs}

Coexistence of different properties of the same spacial anisotropy in one material is crucial for multiferroics. To understand the relations between properties of different types in the same MPG, the structure of MPGs and PRPGs encoded in the group type will be explained and the mapping at the level of those types instead of individual groups will be discussed.

Each of the 122 MPGs can be assigned to one of 6 categories~\cite{Fabrykiewicz2023a}, 12 types~\cite{Kopsky1976,Fabrykiewicz2023a} and 21 extended types introduced below. Recently, similar classification of MPGs into 12 types based on five categories of polarizations and five categories of chirality was published~\cite{Winkler2025}. All 122 MPGs can be derived from 11 proper-rotation-only groups (1, 2, 222, 4, 422, 3, 32, 6, 622, 23 and 432), also called `oriented Laue classes'~\cite{Kopsky2001, Kopsky2006}, by adding antisymmetries as independent operations and/or multiplying half of the operations by them. By multiplying half of the operations of the oriented Laue class by any antisymmetry one can break the spacial equivalence of some directions, e.g. in both $422$ and $4mm$ groups, directions [100] and [110] are equivalent, which is not the case in $\bar{4}2m$ group. This leads to a split of some MPG types from~\cite{Fabrykiewicz2023a} see Table~\ref{Table_multiferroic} and Table~S3 in SM~\cite{SM_joint}, where `iso' and `aniso' superscripts denote whether or not the spacial equivalence of some directions in respect to white groups is broken. Each of the 122 MPGs is uniquely defined by the intersection of (i) oriented Laue class and (ii) extended MPG type.

The 32 CPG and 32 PRPG are collected in Table~\ref{Table_groups}. First column in each row gives the PRPG group-types. 'White' refers to the 11 pure rotation groups only. 'Gray' refers to the 11 groups with the $1^\dagger$ operation as generator. 'B\&W$^\text{iso}$' and 'B\&W$^\text{aniso}$' refer to groups with half of operations multiplied by $1^\dagger$, but where $1^\dagger$ is not a generator (similar to black-and-white MPGs). 'B\&W$^\text{aniso}$' refers to groups in which only a half of rotation axes perpendicular to the main axis is multiplied by $1^\dagger$, i.e. $6^\dagger2^\dagger2$, $4^\dagger2^\dagger2$ and $2^\dagger2^\dagger2$. It means that PRPGs of 'B\&W$^\text{aniso}$' type have higher anisotropy in the plane perpendicular to unique axis than corresponding white, gray and B\&W$^\text{iso}$ PRPGs. White, gray and B\&W PRPGs corresponds to chiral, centrosymmetric and simultaneously non-chiral and non-centrosymmetric non-magnetic point groups as given in Table~\ref{Table_groups}. With this distinction a PRPG is uniquely defined by providing both: oriented Laue class and PRPG type.


\section{Applications}

\subsection{Application to multiferroics}

Traditionally, the symmetry of the physical system is described by one MPG which describes all properties. In our approach the symmetry of the physical system is described by the set of PRPGs, in which each PRPG is obtained for different  physical property type and encodes information about this property. The mapping performed for a single property type cause losing information about other properties types. However, the set of PRPGs encodes the same information as one MPG in the traditional approach.

Table~\ref{Table_multiferroic} summarizes the mapping of MPGs into PRPGs on the level of group types. MPGs belonging to each of the 21 extended types have in principle different multiferroic properties. Subscripts 1, 2 and 3 of a B\&W$^\text{aniso}$ PRPGs type denote a different orientation of the axis (setting) of the same group, so they allow for the same number of $1^\dagger$-odd physical property components. Gray PRPGs do not allow for any component of $1^\dagger$-odd physical property. Depending on the spatially anisotropic character of the considered $1^\dagger$-odd physical property, a given property component is only allowed in one of the white, B\&W$^\text{iso}$, B\&W$^\text{aniso}_1$, B\&W$^\text{aniso}_2$ or B\&W$^\text{aniso}_3$ PRPGs or in none of them. The full-gray MPGs are always non-ferroic, as they allow only for even-type physical properties. The $\hat{I}$-gray and B\&W-$\hat{I}$-gray with $\hat{I}\in\{\red{1},\blue{1},\green{1}\}$ MPG types allow for even- and one of non-even-type of physical property, with the non-even-type being magnetic, electric and magnetoelectric for $\hat{I}=\red{1}$, $\hat{I}=\blue{1}$, and $\hat{I}=\green{1}$, respectively. Other MPG types allow for all four types of physical properties. Due to the spatially anisotropic part of symmetry of the considered $1^\dagger$-odd physical property it is possible that only some or none of white, B\&W$^\text{iso}$, B\&W$^\text{aniso}$ PRPGs types allow for the given property.

For example let us consider the $1^\dagger$-odd vector in PRPGs derived from the $422$ oriented Laue class. Its $x$-, $y$-, and $z$-components have $\infty2^\dagger_x$, $\infty2^\dagger_y$ and $\infty2^\dagger_z$ symmetries, respectively. According to the Neumann's principle~\cite{Neumann_book, Voigt_book, Minnigerode_1884, Curie1884, Shubnikov_Koptsik_book} the $z$-component is only allowed by group $42^\dagger2^\dagger$ and $x$- as well as $y$- components are not allowed by neither of $422$ (white), $42^\dagger2^\dagger$ (B\&W$^\text{iso}$), $4^\dagger22^\dagger$ (B\&W$^\text{aniso}_1$), $4^\dagger2^\dagger2$ (B\&W$^\text{aniso}_2$) and $4221^\dagger$ (gray).  Consulting Table~\ref{Table_multiferroic} with this knowledge, one will find that for 422 class and $1^\dagger$-odd vector only $4mm$ (\blue{1}-B\&W$^\text{iso}$, ferroelectric and ferrotoroidal), $42'2'$ (\red{1}-B\&W$^\text{iso}$, ferromagnetic and ferrotoroidal) and $4m'm'$ (\green{1}-B\&W$^\text{iso}$, ferromagnetic and ferroelectric) MPGs are multiferroic for vectors. The other example is $m'm2'$, which is unique multiferroic MPG being the only member of full-B\&W$^\text{aniso}$ type. It allows for ferromagnetic ($2^\dagger22^\dagger$, $y$-axis direction), ferroelectric ($2^\dagger2^\dagger2$, $z$-axis direction) and ferrotoroidic ($22^\dagger2^\dagger$, $x$-axis direction) moments along three directions perpendicular to each other~\cite{Schmid2008}.

The present approach keeps unchanged spatially anisotropic part of system and property symmetry. Below this general concept will be applied into multipolar degrees of freedom and physical properties tensors.

\begin{table*}
	\begin{ruledtabular}
		\caption{\label{Table_multiferroic} The magnetic point groups (MPG) types mapped into property related point groups (PRPGs) types, showing different multiferroic behavior. All 122 magnetic point groups are given. Explicit realization of the MPGs mapping into PRPGs is provided in Table S1 in Supplemental Material~\cite{SM_joint}.}
		\begin{tabular}{llcccc}
			&                                                                            &                   \multicolumn{4}{c}{Property-related groups}                    \\ \cline{3-6}
			\multicolumn{2}{c}{Magnetic point groups}                                    &  \textcolor{blue}{M}   &   \textcolor{red}{E}   & \textcolor{green}{ME}  & even  \\ \cline{1-2}\cline{3-6}
			full-white                         & $1$, $2$, $222$, $4$, $422$, $3$, $32$, $6$, $622$, $23$, $432$            &         white          &         white          &         white          &       \\ \cline{1-2}\cline{3-5}
			\blue{1}-B\&W$^\text{iso}$         & $m$, $\bar{4}$, $4mm$, $\bar{6}$, $6mm$, $3m$, $\bar{4}3m$                 &         white          &   B\&W$^\text{iso}$    &   B\&W$^\text{iso}$    &       \\
			\red{1}-B\&W$^\text{iso}$          & $2'$, $4'$, $42'2'$, $6$, $62'2'$, $32'$, $4'32'$                          &   B\&W$^\text{iso}$    &         white          &   B\&W$^\text{iso}$    &       \\
			\green{1}-B\&W$^\text{iso}$        & $m'$, $\bar{4}'$, $4m'm'$, $\bar{6}'$, $6m'm'$, $3m'$, $\bar{4}'3m'$       &   B\&W$^\text{iso}$    &   B\&W$^\text{iso}$    &         white          &       \\ \cline{1-2}\cline{3-5}
			\blue{1}-B\&W$^\text{aniso}$       & $mm2$, $\bar{6}m2$, $\bar{4}2m$                                            &         white          & B\&W$^\text{aniso}_1$  & B\&W$^\text{aniso}_1$  &       \\
			\red{1}-B\&W$^\text{aniso}$        & $2'2'2$, $6'22'$, $4'22'$                                                  & B\&W$^\text{aniso}_1$  &         white          & B\&W$^\text{aniso}_1$  &       \\
			\green{1}-B\&W$^\text{aniso}$      & $m'm'2$, $\bar{6}'m'2$, $\bar{4}'2m'$                                      & B\&W$^\text{aniso}_1$  & B\&W$^\text{aniso}_1$  &         white          &       \\ \cline{1-2}\cline{3-5}
			full-B\&W$^\text{\blue{1}-iso}$    & $\bar{6}m'2'$, $\bar{4}2'm'$                                               &   B\&W$^\text{iso}$    & B\&W$^\text{aniso}_1$  & B\&W$^\text{aniso}_2$  &       \\
			full-B\&W$^\text{\red{1}-iso}$     & $6'mm'$, $4'm'm$                                                           & B\&W$^\text{aniso}_1$  &   B\&W$^\text{iso}$    & B\&W$^\text{aniso}_1$  &       \\
			full-B\&W$^\text{\green{1}-iso}$   & $\bar{6}'m2'$, $\bar{4}'2'm$                                               & B\&W$^\text{aniso}_1$  & B\&W$^\text{aniso}_2$  &   B\&W$^\text{iso}$    &       \\ \cline{1-2}\cline{3-5}
			full-B\&W$^\text{aniso}$           & $m'm2'$                                                                    & B\&W$^\text{aniso}_1$  & B\&W$^\text{aniso}_2$  & B\&W$^\text{aniso}_3$  &       \\ \cline{1-2}\cline{3-5}
			\multirow{2}{*}{\blue{1}-gray}     & $\bar{1}$, $2/m$, $mmm$, $4/m$, $4/mmm$,                                   & \multirow{2}{*}{white} & \multirow{2}{*}{gray}  & \multirow{2}{*}{gray}  &       \\
			& $\bar{3}$, $\bar{3}m$, $6/m$, $6/mmm$, $\bar{3}m$, $m\bar{3}m$             &                        &                        &                        & white \\
			\multirow{2}{*}{\red{1}-gray}      & $11'$, $21'$, $2221'$, $41'$, $4221'$,                                     & \multirow{2}{*}{gray}  & \multirow{2}{*}{white} & \multirow{2}{*}{gray}  &       \\
			& $31'$, $321'$, $61'$, $6221'$, $231'$, $4321'$                             &                        &                        &                        &       \\
			\multirow{2}{*}{\green{1}-gray}    & $\bar{1}'$, $2/m'$, $m'm'm'$, $4/m'$, $4/m'm'm'$,                          & \multirow{2}{*}{gray}  & \multirow{2}{*}{gray}  & \multirow{2}{*}{white} &       \\
			& $\bar{3}'$, $\bar{3}m'$, $6/m'$, $6/m'm'm'$, $\bar{3}'m'$, $m'\bar{3}'m'$  &                        &                        &                        &       \\ \cline{1-2}\cline{3-5}
			B\&W$^\text{iso}$-\blue{1}-gray    & $2'/m'$, $4'/m$, $4/mm'm'$, $6'/m'$, $6/mm'm'$, $\bar{3}m'$, $m\bar{3}m'$  &   B\&W$^\text{iso}$    &          gray          &          gray          &       \\
			B\&W$^\text{iso}$-\red{1}-gray     & $m1'$, $\bar{4}1'$, $4mm1'$, $\bar{6}1'$, $6mm1'$, $3m1'$, $\bar{4}3m1'$   &          gray          &   B\&W$^\text{iso}$    &          gray          &       \\
			B\&W$^\text{iso}$-\green{1}-gray   & $2'/m$, $4'/m'$, $4/m'mm$, $6'/m$, $6/m'mm$, $\bar{3}'m$, $m'\bar{3}'m$    &          gray          &          gray          &   B\&W$^\text{iso}$    &       \\ \cline{1-2}\cline{3-5}
			B\&W$^\text{aniso}$-\blue{1}-gray  & $m'm'm$, $4'/mm'm$, $6'/m'mm'$                                             & B\&W$^\text{aniso}_1$  &          gray          &          gray          &       \\
			B\&W$^\text{aniso}$-\red{1}-gray   & $mm21'$, $\bar{4}2m1'$, $\bar{6}m2.1'$                                     &          gray          & B\&W$^\text{aniso}_1$  &          gray          &       \\
			B\&W$^\text{aniso}$-\green{1}-gray & $m'mm$, $4'/m'm'm$, $6'/mmm'$                                              &          gray          &          gray          & B\&W$^\text{aniso}_1$  &       \\ \cline{1-2}\cline{3-5}
			\multirow{2}{*}{full-gray}         & $\bar{1}1'$, $2/m1'$, $mmm1'$, $4/m1'$, $4/mmm1'$,                         & \multirow{2}{*}{gray}  & \multirow{2}{*}{gray}  & \multirow{2}{*}{gray}  &       \\
			& $\bar{3}1'$, $\bar{3}m1'$, $6/m1'$, $6/mmm1'$, $m\bar{3}1'$, $m\bar{3}m1'$ &                        &                        &                        &
		\end{tabular}
	\end{ruledtabular}
\end{table*}


\subsection{Application to multipoles}

Multipole moments form an orthogonal basis for the decomposition of any function of spherical angles in three-dimensional Euclidean space. Basis functions for irreducible representations of SO(3) are complex spherical harmonics. It is useful to use real combinations of spherical harmonics such as tesseral and cubic harmonics, which are compatible with irreducible representations of the hexagonal 6/mmm and cubic m$\bar{3}$m point groups, respectively. Cubic and tesseral harmonics are identical for monopole, dipole and quadrupole but differ for higher order multipoles. The present paper use the definitions of the multipole components from~\cite{Hayami2018JPSoJ, Hayami2018, Yatsushiro2021, Yatsushiro2022} and multipole types from Table~\ref{Table:types}.

To focus on the spatially anisotropic part of the multipoles symmetry one can consider abstract $1^\dagger$-odd multipoles denoted by the symbol $X$ with indices indicating particular components. The symmetry of each component of multipoles up to rank 4 is given in Table~S4 in SM~\cite{SM_joint}. Let \{G\} denote a set of all symmetry allowed multipole components by the group G. In Fig.~\ref{Fig_graph} \{G\} corresponds to $F(G)$. A multipole component belongs to \{G\} when G is a subgroup of a multipole component symmetry group or equivalently when it belongs to the totally symmetric irreducible representation of the PRPG~\cite{Hayami2018, Yatsushiro2021, Yatsushiro2022}. Table~\ref{Table:active_multipoles} classifies all active $1^\dagger$-odd multipoles up to rank 4 in property-related groups, i.e. each row contains a subset of \{G\} with the PRPG G being given in the first column. Table~\ref{Table:active_multipoles} contains the same information as (approximately in total four full journal pages long) Tables V-XI from~\cite{Yatsushiro2021}, where active multipole components of each of 4 types for each of 122 magnetic point groups are given explicitly. An extended version of Table~\ref{Table:active_multipoles} constitutes Table~S4 in SM~\cite{SM_joint}. Symmetry allowed multipole components depend on the group setting (orientation of axes). However, the total number of components is setting-invariant as given in Table~\ref{Table:decomposition}.


\subsection{Application to tensors}

\subsubsection{Tensors intrinsic symmetry encoded in modified Jahn symbol}

Tensors, similar to multipoles, can be classified into four types regarding their behavior under action of \blue{1}, \red{1} and \green{1} antisymmetries, as given in Table~\ref{Table:types}. Tensor type together with rank and information about symmetrization and antisymmetrization of indices consist of tensor intrinsic symmetry encoded in the Jahn symbol~\cite{Jahn_notation, MTENSOR}. Jahn symbols are build around polar time-even vectors $V$ (i.e. electric-type vectors), where each $V$ corresponds to one of the tensor indices. Therefore, the number of $V$ corresponds to the tensor rank. Symmetrization and antisymmetrization of indices is denoted by including $V$s corresponding to (anti)symmetrized indices in square and curly brackets, respectively. Consequent $V$s which are not separated by brackets are collected into $V^N$ symbol, where $N$ denote the number of consequent tensor indices. Since $V$ is polar (i.e. \blue{1}-odd), Jahn symbols described up to now represent only even-type even-rank and electric-type odd-rank tensors, see Table~\ref{Table:types}. To cover other combinations of tensor ranks and types, $V$s in Jahn symbol may be preceded by $a$, $e$ or $ae$ corresponding to electric-, magnetic- and magnetoelectric-type monopoles, respectively. Jahn symbol of some transport and nonlinear optical susceptibility tensors could also contain asterisk~\cite{MTENSOR} which denote different tensor-type of symmetric and antisymmetric part of the tensor~\cite{Grimmer2017} or situation where time reversal connect given tensor property with another one~\cite{MTENSOR}.

\pagebreak

\begin{table}[h!]
	\begin{ruledtabular}
		\caption{\label{Table:active_multipoles} Active $1^\dagger$-odd multipoles up to rank 4 in property related groups. An extended version of this table is included as Table S4 in Supplemental Material (SM)~\cite{SM_joint}. Groups containing $1^\dagger$ as generator, i.e. $11^\dagger$, $21^\dagger$, $31^\dagger$, $41^\dagger$, $61^\dagger$, $2221^\dagger$, $3211^\dagger$, $4221^\dagger$, $6221^\dagger$, $231^\dagger$, $4321^\dagger$, are omitted as in them no multipoles are active. Note that each property-related group corresponds to a set of magnetic point groups depending on the type of multipole (Table~\ref{Table:types}).}
		\begin{adjustbox}{max width=1.0\hsize}
			\begin{tabular}{lccccc}
				$l=$                  &  $0$  &  $1$  &       $2$       &               $3$                &                                                                                                                                                                         $4$                                                                                                                                                                         \\ \hline
				$1$                   & \multicolumn{5}{l}{All multipole components are active.}                                                                                                                                                                                                                                                                                                                                                                 \\
				$2_c$                 & \multicolumn{5}{l}{Active multipole components in the footnote\footnote{In $2_c$ active are: $X_0$ monopole, $X_z$ dipole, $X_u$, $X_v$, $X_{xy}$ quadrupole, $X_{xyz}$, $X_z^\alpha$, $X_z^\beta$ octupole, and $X_4$, $X_{4u}$, $X_{4v}$, $X_{4z}^\alpha$, $X_{4z}^\beta$ ($X_{40}$, $X_{4u}^{\beta1}$, $X_{4v}^{\beta1}$, $X_{4u}^{\beta2}$, $X_{4v}^{\beta2}$) hexadecapole components.}.}                           \\
				$2_c^\dagger$         & \multicolumn{5}{l}{Active multipole components in the footnote\footnote{In $2_c^\dagger$ active are: $X_x$, $X_y$ dipole, $X_{yz}$, $X_{zx}$ quadrupole, $X_x^\alpha$, $X_y^\alpha$, $X_x^\beta$, $X_y^\beta$ ($X_{3a}$, $X_{3b}$, $X_{3u}$, $X_{3v}$) octupole, and $X_{4x}^\alpha$, $X_{4y}^\alpha$, $X_{4x}^\beta$, $X_{4y}^\beta$ ($X_{4a}$, $X_{4b}$, $X_{4u}^\alpha$, $X_{4v}^\alpha$) hexadecapole components.}.} \\
				$222$                 & $X_0$ &       &  $X_u$, $X_v$   &            $X_{xyz}$             &                                                                                                                                                              $X_4$, $X_{4u}$, $X_{4v}$                                                                                                                                                              \\
				$2^\dagger2^\dagger2$ &       & $X_z$ &    $X_{xy}$     &    $X_z^\alpha$, $X_z^\beta$     &                                                                                                                                                           $X_{4z}^\alpha$, $X_{4z}^\beta$                                                                                                                                                           \\
				$4$                   & $X_0$ & $X_z$ &      $X_u$      &           $X_z^\alpha$           &                                                                                                                                                          $X_4$, $X_{4u}$, $X_{4z}^\alpha$                                                                                                                                                           \\
				$4^\dagger$           &       &       & $X_v$, $X_{xy}$ &      $X_{xyz}$, $X_z^\beta$      &                                                                                                                                                              $X_{4v}$, $X_{4z}^\beta$                                                                                                                                                               \\
				$422$                 & $X_0$ &       &      $X_u$      &                                  &                                                                                                                                                                   $X_4$, $X_{4u}$                                                                                                                                                                   \\
				$42^\dagger2^\dagger$ &       & $X_z$ &                 &           $X_z^\alpha$           &                                                                                                                                                                   $X_{4z}^\alpha$                                                                                                                                                                   \\
				$4^\dagger22^\dagger$ &       &       &      $X_v$      &            $X_{xyz}$             &                                                                                                                                                                      $X_{4v}$                                                                                                                                                                       \\
				$23$                  & $X_0$ &       &                 &            $X_{xyz}$             &                                                                                                                                                                        $X_4$                                                                                                                                                                        \\
				$432$                 & $X_0$ &       &                 &                                  &                                                                                                                                                                        $X_4$                                                                                                                                                                        \\
				$4^\dagger32^\dagger$ &       &       &                 &            $X_{xyz}$             &                                                                                                                                                                                                                                                                                                                                                     \\
				$3$                   & $X_0$ & $X_z$ &      $X_u$      & $X_z^\alpha$, $X_{3a}$, $X_{3b}$ &                                                                                                                                                            $X_{40}$, $X_{4a}$, $X_{4b}$                                                                                                                                                             \\
				$321$                 & $X_0$ &       &      $X_u$      &             $X_{3a}$             &                                                                                                                                                                 $X_{40}$, $X_{4a}$                                                                                                                                                                  \\
				$32^\dagger1$         &       & $X_z$ &                 &      $X_z^\alpha$, $X_{3b}$      &                                                                                                                                                                      $X_{4b}$                                                                                                                                                                       \\
				$6$                   & $X_0$ & $X_z$ &      $X_u$      &           $X_z^\alpha$           &                                                                                                                                                                      $X_{40}$                                                                                                                                                                       \\
				$6^\dagger$           &       &       &                 &        $X_{3a}$, $X_{3b}$        &                                                                                                                                                                 $X_{4a}$, $X_{4b}$                                                                                                                                                                  \\
				$622$                 & $X_0$ &       &      $X_u$      &                                  &                                                                                                                                                                      $X_{40}$                                                                                                                                                                       \\
				$62^\dagger2^\dagger$ &       & $X_z$ &                 &           $X_z^\alpha$           &                                                                                                                                                                                                                                                                                                                                                     \\
				$6^\dagger22^\dagger$ &       &       &                 &             $X_{3a}$             &                                                                                                                                                                      $X_{4a}$
			\end{tabular}%
		\end{adjustbox}
	\end{ruledtabular}
\end{table}

\begin{table}[h!]
	\begin{ruledtabular}
		\caption{\label{Table:decomposition} Upper part: decomposition of tensors up to rank 4 into multipoles. Lower part: number of symmetry allowed components for multipoles and tensors up to rank 4 in property related groups. Monopole, dipole, quadrupole, octupole and hexadecapole are denoted by M, D, Q, O and H, respectively. The "-" symbol stands for zero, and is used to highlight groups which do not allow for an effect represented by given tensors. Number of allowed components is independent from the group setting. Each multipole and tensor is $1^\dagger$-odd. Note that each property-related group corresponds to set of magnetic point groups depending on the type of multipole (Table~\ref{Table:types}).}
		\begin{tabular}{lccccccccccccccccc}
			\rotatebox[origin=c]{90}{Tensors}                     & \rotatebox[origin=c]{90}{$x$} & \rotatebox[origin=c]{90}{$xG$, $x\{G^2\}$} &   &   &   & \rotatebox[origin=c]{90}{List in the footnote\footnote{List: $xG^2$, $xG\{G^2\}$, $x\{G^3\}$, $x\{G^2\}\{G^2\}$, $x\{G^4\}$, $x[\{G^2\}\{G^2\}]$}.} & \rotatebox[origin=c]{90}{$x[G^2]$} & \rotatebox[origin=c]{90}{$xG^3$, $xG^2\{G^2\}$, $xG\{G^3\}$} & \rotatebox[origin=c]{90}{$xG[G^2]$, $x\{G^2\}[G^2]$} & \rotatebox[origin=c]{90}{$x[G^3]$} & \rotatebox[origin=c]{90}{$xG^4$} & \rotatebox[origin=c]{90}{$xG^2[G^2]$} & \rotatebox[origin=c]{90}{$xG[G^3]$} & \rotatebox[origin=c]{90}{$x[G^2][G^2]$} & \rotatebox[origin=c]{90}{$x[[G^2][G^2]]$} & \rotatebox[origin=c]{90}{$x[G^4]$} & \rotatebox[origin=c]{90}{$x\{[G^2][G^2]\}$} \\ \hline
			\multirow{5}{*}{\rotatebox[origin=c]{90}{Multipoles}} &               M               &                                            &   &   &   &                                                                          1                                                                          &                 1                  &                              1                               &                          -                           &                 -                  &                3                 &                   2                   &                  1                  &                    2                    &                     2                     &                 1                  &                      -                      \\
			                                                      &                               &                     D                      &   &   &   &                                                                          1                                                                          &                 -                  &                              3                               &                          2                           &                 1                  &                6                 &                   3                   &                  1                  &                    1                    &                     -                     &                 -                  &                      1                      \\
			                                                      &                               &                                            & Q &   &   &                                                                          1                                                                          &                 1                  &                              2                               &                          1                           &                 -                  &                6                 &                   4                   &                  2                  &                    3                    &                     2                     &                 1                  &                      1                      \\
			                                                      &                               &                                            &   & O &   &                                                                          -                                                                          &                 -                  &                              1                               &                          1                           &                 1                  &                3                 &                   2                   &                  1                  &                    1                    &                     -                     &                 -                  &                      1                      \\
			                                                      &                               &                                            &   &   & H &                                                                          -                                                                          &                 -                  &                              -                               &                          -                           &                 -                  &                1                 &                   1                   &                  1                  &                    1                    &                     1                     &                 1                  &                      -                      \\ \hline
			$1$                                                   &               1               &                     3                      & 5 & 7 & 9 &                                                                          9                                                                          &                 6                  &                              27                              &                          18                          &                 10                 &                81                &                  54                   &                 30                  &                   36                    &                    21                     &                 15                 &                     15                      \\
			$2$                                                   &               1               &                     1                      & 3 & 3 & 5 &                                                                          5                                                                          &                 4                  &                              13                              &                          8                           &                 4                  &                41                &                  28                   &                 16                  &                   20                    &                    13                     &                 9                  &                      7                      \\
			$2^\dagger$                                           &               -               &                     2                      & 2 & 4 & 4 &                                                                          4                                                                          &                 2                  &                              14                              &                          10                          &                 6                  &                40                &                  26                   &                 14                  &                   16                    &                     8                     &                 6                  &                      8                      \\
			$222$                                                 &               1               &                     -                      & 2 & 1 & 3 &                                                                          3                                                                          &                 3                  &                              6                               &                          3                           &                 1                  &                21                &                  15                   &                  9                  &                   12                    &                     9                     &                 6                  &                      3                      \\
			$2^\dagger2^\dagger2$                                 &               -               &                     1                      & 1 & 2 & 2 &                                                                          2                                                                          &                 1                  &                              7                               &                          5                           &                 3                  &                20                &                  13                   &                  7                  &                    8                    &                     4                     &                 3                  &                      4                      \\
			$4$                                                   &               1               &                     1                      & 1 & 1 & 3 &                                                                          3                                                                          &                 2                  &                              7                               &                          4                           &                 2                  &                21                &                  14                   &                  8                  &                   10                    &                     7                     &                 5                  &                      3                      \\
			$4^\dagger$                                           &               -               &                     -                      & 2 & 2 & 2 &                                                                          2                                                                          &                 2                  &                              6                               &                          4                           &                 2                  &                20                &                  14                   &                  8                  &                   10                    &                     6                     &                 4                  &                      4                      \\
			$422$                                                 &               1               &                     -                      & 1 & - & 2 &                                                                          2                                                                          &                 2                  &                              3                               &                          1                           &                 -                  &                11                &                   8                   &                  5                  &                    7                    &                     6                     &                 4                  &                      1                      \\
			$42^\dagger2^\dagger$                                 &               -               &                     1                      & - & 1 & 1 &                                                                          1                                                                          &                 -                  &                              4                               &                          3                           &                 2                  &                10                &                   6                   &                  3                  &                    3                    &                     1                     &                 1                  &                      2                      \\
			$4^\dagger22^\dagger$                                 &               -               &                     -                      & 1 & 1 & 1 &                                                                          1                                                                          &                 1                  &                              3                               &                          2                           &                 1                  &                10                &                   7                   &                  4                  &                    5                    &                     3                     &                 2                  &                      2                      \\
			$23$                                                  &               1               &                     -                      & - & 1 & 1 &                                                                          1                                                                          &                 1                  &                              2                               &                          1                           &                 1                  &                7                 &                   5                   &                  3                  &                    4                    &                     3                     &                 2                  &                      1                      \\
			$432$                                                 &               1               &                     -                      & - & - & 1 &                                                                          1                                                                          &                 1                  &                              1                               &                          -                           &                 -                  &                4                 &                   3                   &                  2                  &                    3                    &                     3                     &                 2                  &                      -                      \\
			$4^\dagger32^\dagger$                                 &               -               &                     -                      & - & 1 & - &                                                                          -                                                                          &                 -                  &                              1                               &                          1                           &                 1                  &                3                 &                   2                   &                  1                  &                    1                    &                     -                     &                 -                  &                      1                      \\
			$3$                                                   &               1               &                     1                      & 1 & 3 & 3 &                                                                          3                                                                          &                 2                  &                              9                               &                          6                           &                 4                  &                27                &                  18                   &                 10                  &                   12                    &                     7                     &                 5                  &                      5                      \\
			$32$                                                  &               1               &                     -                      & 1 & 1 & 2 &                                                                          2                                                                          &                 2                  &                              4                               &                          2                           &                 1                  &                14                &                  10                   &                  6                  &                    8                    &                     6                     &                 4                  &                      2                      \\
			$32^\dagger$                                          &               -               &                     1                      & - & 2 & 1 &                                                                          1                                                                          &                 -                  &                              5                               &                          4                           &                 3                  &                13                &                   8                   &                  4                  &                    4                    &                     1                     &                 1                  &                      3                      \\
			$6$                                                   &               1               &                     1                      & 1 & 1 & 1 &                                                                          3                                                                          &                 2                  &                              7                               &                          4                           &                 2                  &                19                &                  12                   &                  6                  &                    8                    &                     5                     &                 3                  &                      3                      \\
			$6^\dagger$                                           &               -               &                     -                      & - & 2 & 2 &                                                                          -                                                                          &                 -                  &                              2                               &                          2                           &                 2                  &                8                 &                   6                   &                  4                  &                    4                    &                     2                     &                 2                  &                      2                      \\
			$622$                                                 &               1               &                     -                      & 1 & - & 1 &                                                                          2                                                                          &                 2                  &                              3                               &                          1                           &                 -                  &                10                &                   7                   &                  4                  &                    6                    &                     5                     &                 3                  &                      1                      \\
			$62^\dagger2^\dagger$                                 &               -               &                     1                      & - & 1 & - &                                                                          1                                                                          &                 -                  &                              4                               &                          3                           &                 2                  &                9                 &                   5                   &                  2                  &                    2                    &                     -                     &                 -                  &                      2                      \\
			$6^\dagger22^\dagger$                                 &               -               &                     -                      & - & 1 & 1 &                                                                          -                                                                          &                 -                  &                              1                               &                          1                           &                 1                  &                4                 &                   3                   &                  2                  &                    2                    &                     1                     &                 1                  &                      1
		\end{tabular}
	\end{ruledtabular}
\end{table}

\pagebreak
\textcolor{white}{.}
\pagebreak
\textcolor{white}{.}

The modified Jahn symbols based on the even-type vector $G$ (electric toroidal moment~\cite{Hlinka2014}) instead of the electric-type vector~$V$ will be used, see Table~\ref{Table:intrinsic_symmetry}. The advantage of this approach is that the tensor type does not depend on the total exponent of $G$ and is fully encoded in the scalar at the beginning of the symbol. Here, a deep warning have to be expressed because G is used for both the group and the even-type vector. For historical reasons both notations are well grounded and he hope that the readers will not confuse them. To focus on the spatially anisotropic part of the tensor symmetry one consider abstract $1^\dagger$-odd tensors, whose modified Jahn symbols begin with the $1^\dagger$-odd monopole $x$. Table S5 in SM~\cite{SM_joint} lists selected tensors~\cite{MTENSOR} with their (modified) Jahn symbol.

\begin{table}
	\caption{\label{Table:intrinsic_symmetry}
	Comparison of tensors intrinsic symmetry symbols. Traditional~\cite{Jahn_notation, MTENSOR} notation mix types between even and odd tensor ranks $N$. Notation used in this paper is independent from tensor rank. Following relation occur: $V=\epsilon G$.}
	\begin{ruledtabular}
		\begin{tabular}{rllllrrr}
			                       &     &           &      &            &                       \multicolumn{3}{c}{\textrm{Tensors}}                       \\ \cline{6-8}
			          \multicolumn{5}{c}{\textrm{Definition}}            & \multicolumn{2}{c}{\textrm{Jahn~\cite{Jahn_notation, MTENSOR}}} &  \textrm{This} \\ \cline{1-5}\cline{6-7}
			         \textrm{Type} & 1   & $\bar{1}$ & $1'$ & $\bar{1}'$ & \text{even }N &                                    \text{odd }N & \textrm{paper} \\ \hline
			           \text{Even} & $+$ & $+$       & $+$  & $+$        &         $V^N$ &                                          $eV^N$ &          $G^N$ \\
			       \text{Magnetic} & $+$ & $+$       & $-$  & $-$        &        $aV^N$ &                                         $aeV^N$ &         $aG^N$ \\
			       \text{Electric} & $+$ & $-$       & $+$  & $-$        &        $eV^N$ &                                           $V^N$ &         $eG^N$ \\
			\text{Magnetoelectric} & $+$ & $-$       & $-$  & $+$        &       $aeV^N$ &                                          $aV^N$ &        $aeG^N$
		\end{tabular}
	\end{ruledtabular}
\end{table}


\subsubsection{Correspondence between tensor components and multipoles}

Tensors can be expressed in the basis of multipole degrees of freedom~\cite{Hayami2018, Yatsushiro2021, Yatsushiro2022}. Each tensor corresponds to a set of multipoles of the same type as the tensor, as shown for tensors up to rank 4 in the upper part of Table~\ref{Table:decomposition}. In this set, two or more independent multipoles of the same rank may appear. The total number of components of those multipoles is equal to the number of tensor components. Each component of a tensor is a linear combination of components of multipoles from the above mentioned set, similar to hybridization of atomic orbitals in valence bond theory~\cite{Pauling1931, Pauling_book}. The combination of correspondence between tensors and multipoles (upper part of Table~\ref{Table:decomposition}) with the number of active multipole components in property-related groups (left part of Table~\ref{Table:decomposition}) leads to the number of non-zero tensor components in property-related groups (bottom right part of Table~\ref{Table:decomposition}).

The scalar ($x$ tensor) and vector ($xG$ tensor) are equivalent to the monopole, $(X_0)$, and the dipole $(X_x, X_y, X_z)$, respectively. The $xG^2$ tensor corresponds to the monopole, dipole and quadrupole as given explicitly in Appendix. The $xG[G^2]$ tensor corresponds to two independent dipoles, one quadrupole and one octupole as given explicitly in the Appendix in cubic harmonics. The $xG[G^2]$ tensor in tesseral harmonics as well as $x[G^3]$, $xG[G^3]$, $x[G^2][G^2]$ tensors expressed in both cubic and tesseral harmonics can be found in~\cite{Yatsushiro2021, Yatsushiro2022}.


\subsection{Tensor form algebra}

\subsubsection{Tensor form}

Tensor form is understood as a set of symmetry allowed tensor components with a set of linear equations between them, which results from the symmetry of the medium.

Classification of forms of over one hundred equilibrium, transport, optical and nonlinear optical susceptibility tensors forms for all 122 MPGs are available online via MTENSOR tool on the Bilbao Crystallographic Server~\cite{MTENSOR, MTENSORatBCS}. Correspondence between original and modified Jahn symbols is given in Table S5 in SM~\cite{SM_joint}. A tensor form for any point group can be easily obtained when the tensor is expressed in terms of multipoles, by checking which multipole components are allowed by given group. Since \{G\} denotes a set of all symmetry allowed multipole components by group G, each tensor form corresponds to a subset of \{G\}. This explains why there are 21 tensor forms for non-even-type tensors and 11 tensor forms for even-type tensors~\cite{Birss1963, Birss_book}, as there are 21 physically-relevant property-related groups among which 11 groups consist of proper rotations only. Therefore, the symbol \{G\}$^\text{J}$ will be used to denote tensor forms of tensors with intrinsic symmetry given by a modified Jahn symbol J in group G. The proposed notation of tensor forms (\{G\}$^\text{J}$ symbols) makes (six full book pages long) Table~7 from~\cite{Birss_book} redundant, as this table gives tensor forms as an intersection of magnetic point group and tensor-type, both covered in the proposed symbols. Moreover, those new symbols carry information about the MPG setting, which is lost in the notation from~\cite{Birss_book}. Comparison of \{G\}$^\text{J}$ symbols with literature notations~\cite{Birss_book, Grimmer1991} is given in Table S6 in SM~\cite{SM_joint}. All relations between sets of multipoles \{G$_i$\} corresponding to different groups G$_i$ naturally transfer to tensor forms \{G$_i$\}$^\text{J}$ with any Jahn symbol J.


\subsubsection{Tensor form algebra}

Now one can extrapolate the relation between groups $\updownarrow$ to the relation between symmetry allowed properties $\raisebox{0.3ex}{\rotatebox[origin=c]{-90}{\scalebox{0.55}{$\longdashleftrightarrow$}}}$ as drawn in a symbolic way in in Fig.~\ref{Fig_graph}.

The group G and the set \{G\} are somehow dual to each other. The maximal group, SO(3)$\times 1^\dagger$, does not allow for any multipole, therefore $\{$SO(3)$\times 1^\dagger\}=\varnothing$ is an empty set. Let F be a subgroup of a group G, then for a number of elements (order) ord(G)~$\geq$~ord(F) occurs and \{G\} is a subset of \{F\}. The minimal, trivial group G~$=1$ does not give any restrictions on multipoles, therefore \{1\} is a set of all multipole components.

\begin{figure}
	\includegraphics[width=1.0\hsize]{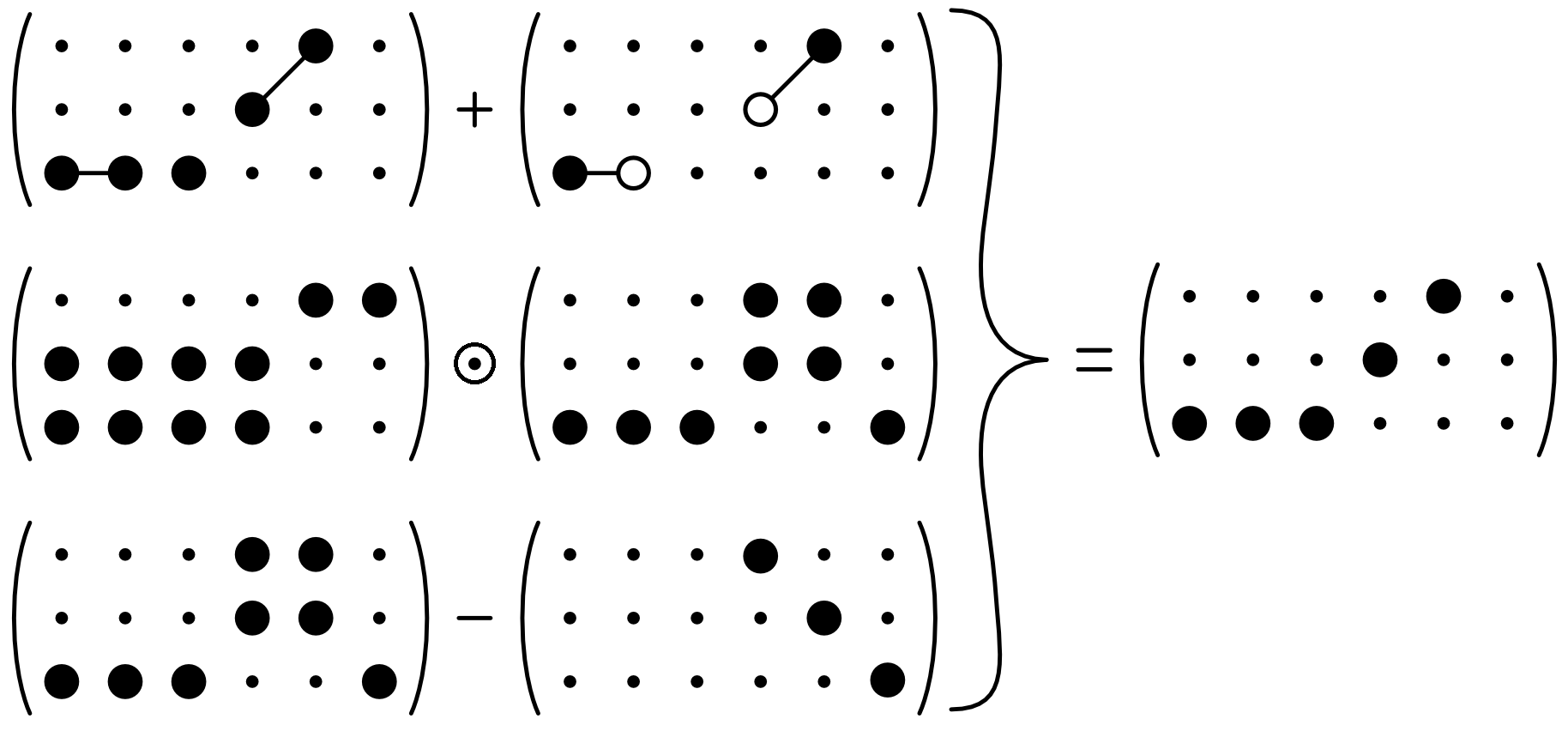} 
	\caption{\label{Fig_example} Example of a $\{2^\dagger2^\dagger2\}^\text{J}$ decomposition into $\{42^\dagger2^\dagger\}^\text{J}+\{4^\dagger2^\dagger2\}^\text{J}$, $\{2_a^\dagger\}^\text{J} \odot \{2_c\}^\text{J}$ and $\{2_c\}^\text{J}-\{222\}^\text{J}$. Iconographic for $\text{J}=x\text{G}[\text{G}]^2$ tensors. Forms are obtained by expressing the $x\text{G}[\text{G}]^2$ tensor in terms of multipoles (Eq.~\ref{Eq_xG[G2]} in Appendix) followed by checking of active multipoles in specific groups (Table~\ref{Table:active_multipoles}). Abbreviated notation $t_{ijk}\rightarrow t_{i\alpha}$, with $jk\rightarrow\alpha$: $11\rightarrow1$; $22\rightarrow2$; $33\rightarrow3$; $23,32\rightarrow4$; $13,31\rightarrow5$ and $12,21\rightarrow6$. Nye's notation~\cite{Nye_book} is used: (i) filled circles and dots denote allowed and forbidden components, respectively, (ii) two connected filled circles denote two equal components, (iii) connected empty and filled circles denote two components with equal absolute value, but opposite sign.}
\end{figure}

Not every set of multipoles corresponds to a group. Therefore, not all expressions involving sets of multipoles and operations from algebra of sets can be replaced by \{G\} with some group G. In the following paragraphs those expressions will be examined for which it is allowed. Fig.~\ref{Fig_example} show an example of how $\{2^\dagger2^\dagger2\}^\text{J}$ can be expressed by different sets with iconographics~\cite{Nye_book} for $\text{J}=x\text{G}[\text{G}]^2$ tensors, e.g. piezomagnetic, piezoelectric and piezotoroidic tensors (all direct effects) following Table S7 in SM~\cite{SM_joint}.

Instead of binary relations $\updownarrow$ between two groups $G_1$ and $G_2$ in Fig.~\ref{Fig_graph}, from now the relations between three groups A, B and C with their symmetry constrained forms \{A\}, \{B\} and \{C\} will be considered, respectively.

\begin{figure}[h!]
	\includegraphics[width=1.0\hsize]{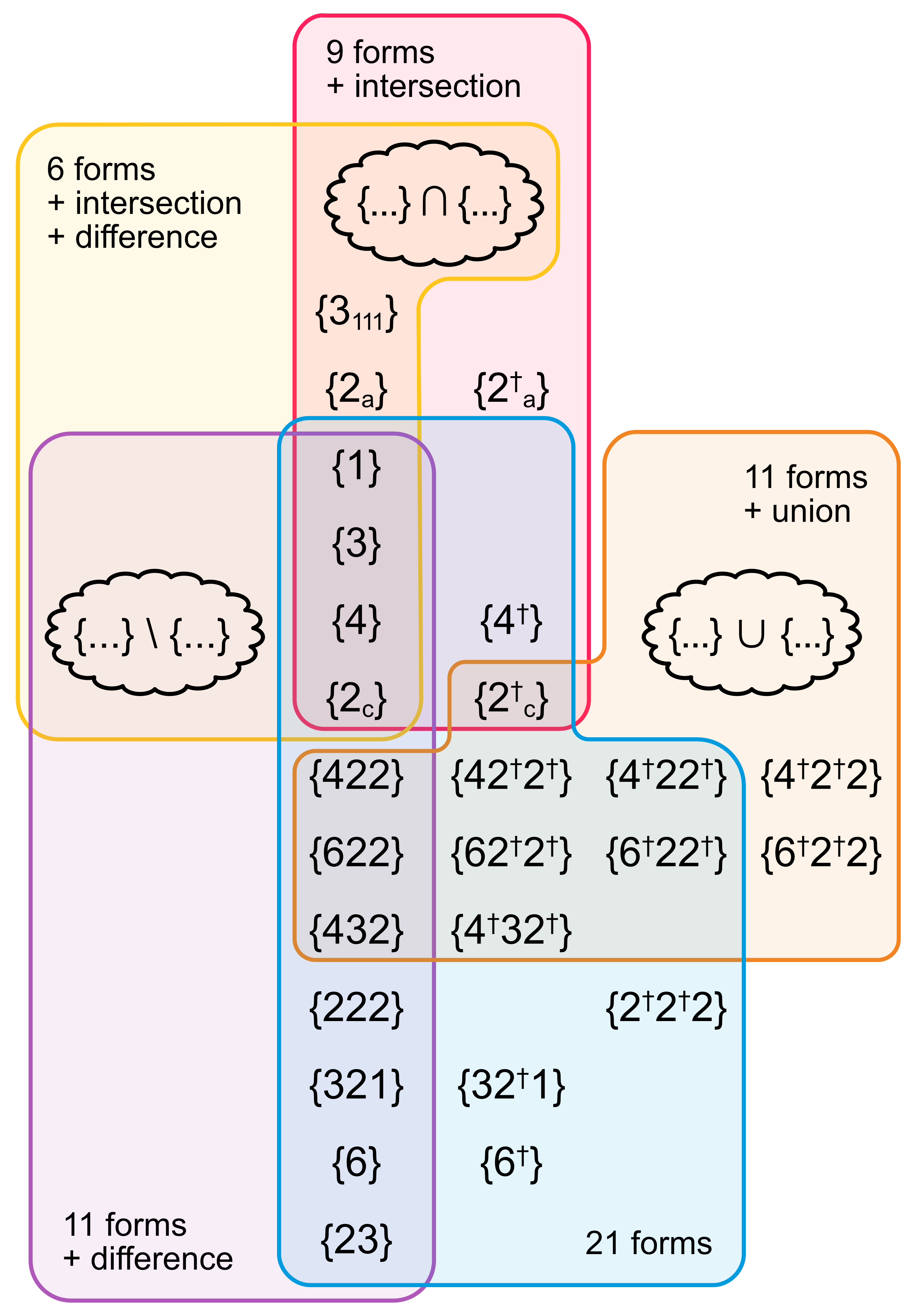} 
	\caption{\label{Fig_sets} Five sets of tensor forms sufficient to describe the tensors of any type of symmetry for all 122 magnetic point groups for given tensor rank. Bubbles denote intersection, union and difference operations. Each row contains groups based on the same proper-rotations-only group (white groups, included in the first column), but have different assignments of daggers. The second column contains B\&W$^\text{iso}$ groups, which have the same symmetry equivalence of principle direction as proper-rotations-only group they are based on. The third and fourth columns contains B\&W$^\text{aniso}$ groups in two different settings for which dagger assignment breaks symmetry equivalence of principle direction in comparison to the proper-rotations-only group they are based on. Gray groups, which contain $1^\dagger$ as an independent operator are not listed as they does not allow for any tensor component.}
\end{figure}

The intersection \{A\}$\cap$\{B\} is equal to \{C\} with C being a minimal common supergroup of groups A and B, e.g. $\{2_a^\dagger\}\cap\{2_c\}=\{2^\dagger2^\dagger2\}$. Any set \{G\} can be non-uniquely decomposed into the intersection of sets corresponding to the cyclic groups generated by each generator of G alone, e.g. $\{622\}=\{3\}\cap\{2_c\}\cap\{2_a\}$. 

The union \{A\}$\cup$\{B\} is equal to \{C\}, when A and B groups share the same oriented Laue class (i.e. come from the same proper rotation group), but have a different assignment of daggers. This ensures that \{A\} and \{B\} sets are disjoint, i.e. $\{\text{A}\}\cap\{\text{B}\}=\varnothing$ and the number of elements of \{C\} is equal to the sum of the number of elements of \{A\} and \{B\}. Then, the C group becomes a maximal common subgroup of A and B, e.g. $\{4^\dagger2^\dagger2\}\cup\{42^\dagger2^\dagger\}=\{2^\dagger2^\dagger2\}$.

The difference \{A\}$\backslash$\{B\} is equal to \{C\}, when ord(A)~$=2$~ord(B) and A is a subgroup of B. The latter ensures that there is no single multipole present in \{B\}, but absent in \{A\} and that the number of elements of \{A\} is equal to the sum of number of elements of sets \{B\} and \{C\}. Then, the group C is obtained by modifying group B by the multiplication by $1^\dagger$ all operations which are present in B but are absent in A, i.e. a similar procedure how magnetic point groups are obtained from crystallographic point groups, e.g. $\{2_c\}\backslash\{222\}=\{2^\dagger2^\dagger2\}$.

The concept of a forms difference appeared in~\cite{Grimmer1991} where other notation of tensor forms was used. Present notation i.e. \{G\} with G being one of PRPGs highlights connection between $\updownarrow$ and $\raisebox{0.3ex}{\rotatebox[origin=c]{-90}{\scalebox{0.55}{$\longdashleftrightarrow$}}}$ in Fig.~\ref{Fig_graph}. Present notation is compared with notation from~\cite{Grimmer1991} in Table S6 in SM~\cite{SM_joint}.

\pagebreak

The multipole sets and tensor forms for 21 physically-important property-related groups in standard setting can be generated using: (i) 9 sets/forms with intersections between them, (ii) 11 sets/forms with unions between them, (iii) 11 sets/forms with differences between them~\cite{Grimmer1991} and (iv) 6 sets/forms with both intersections and differences between them as given in Venn diagram in Fig.~\ref{Fig_sets} and explicit equations in Table S7 in SM~\cite{SM_joint}.


\subsubsection{Element-wise vector algebra instead of algebra of sets}

Instead of the algebra of sets one can use vector algebra. To do so, one can replace each set with a vector. Each coordinate of the vector corresponds to one multipole component. The vector coordinate values indicate whether a given multipole component is permitted (a value of +1) or not (a value of 0). Intersection ($\cap$), union ($\cup$) and difference ($\backslash$) of sets are replaced with element-wise product ($\odot$), sum ($+$) and difference ($-$) of vectors, respectively as explained in Table~\ref{Table:Set_operations}. There are two exceptions marked with brackets, i.e. $1+1=2$ and $0-1=-1$, but they have no correspondence with the multipole sets. Vector algebra can simplify some expressions between sets/forms.

To show the benefit of using vector algebra over using algebra of sets, let us express $\{2_b\}$ in terms of $\{2_a\}$, $\{2_c\}$ and $\{1\}$:
\begin{eqnarray}
	&& \{2_b\} = \{1\}\backslash\{2_b^\dagger\} = \{1\}\backslash\bigg[\{2^\dagger2^\dagger2\} \cup \{22^\dagger2^\dagger\}\bigg] \nonumber\\
	&& = \{1\}\backslash\bigg[\bigg(\{2_a\}\backslash\{222\}\bigg)\cup\bigg(\{2_c\}\backslash\{222\}\bigg)\bigg] \nonumber\\
	&& = \{1\}\backslash\bigg[\bigg(\{2_a\}\backslash(\{2_a\}\cap\{2_c\})\bigg)\cup\bigg(\{2_c\}\backslash(\{2_a\}\cap\{2_c\})\bigg)\bigg] \nonumber\\
	&& //\text{algebra of sets} \rightarrow \text{vector algebra}// \nonumber\\
	&& = \{1\} - \bigg[\bigg(\{2_a\}-(\{2_a\} \{2_c\})\bigg)+\bigg(\{2_c\}-(\{2_a\} \{2_c\})\bigg)\bigg] \nonumber\\
	&& = \{1\}+2\{2_a\}\{2_c\}-\{2_a\}-\{2_c\}.
\end{eqnarray}
In the intermediate steps $\{2_b^\dagger\}$, $\{2^\dagger2^\dagger2\}$, $\{22^\dagger2^\dagger\}$ and $\{222\}$ is used. One only need to know that they exist, but one do not need to know their explicit form. Other examples are given in Table~S7 in SM~\cite{SM_joint}. 

\begin{table}
	\caption{\label{Table:Set_operations} Comparison of intersection ($\cap$), union ($\cup$) and difference ($\backslash$) in algebra of sets (upper part) with element-wise product ($\odot$), sum ($+$) and difference ($-$) in vector algebra (lower part). Each line corresponds to one representative multipole element. In the upper part, the "$\cdot$" and "$\bullet$" denote that the multipole component is absent or present in the set, respectively. In the lower part values of "0" and "1" denote coordinates of vector. There is a one-to-one correspondence between algebra of sets and vector algebra, when exchanging "$\cdot$" and "$\bullet$" to "0" and "1", respectively, everywhere apart from two places marked by brackets. Requirements for expressing difference and union of sets as a set corresponding to single group\footnote{For difference it is enough that A is a subgroup of B and for union $\{A\}\cap\{B\}=\varnothing$ is needed.} ensures that operations leading to outcome included in brackets are not realized.}
	\begin{ruledtabular}
		\begin{tabular}{ccccc}
			\{A\}   &   \{B\}   & $\{\text{A}\}\cap\{\text{B}\}$  & $\{\text{A}\}\cup\{\text{B}\}$ & $\{\text{A}\}\backslash\{\text{B}\}$ \\ \hline
			$\cdot$  &  $\cdot$  &             $\cdot$             &            $\cdot$             &               $\cdot$                \\
			$\cdot$  & $\bullet$ &             $\cdot$             &           $\bullet$            &              $(\cdot)$               \\
			$\bullet$ &  $\cdot$  &             $\cdot$             &           $\bullet$            &              $\bullet$               \\
			$\bullet$ & $\bullet$ &            $\bullet$            &          $(\bullet)$           &               $\cdot$                \\ \hline\hline
			\{A\}   &   \{B\}   & $\{\text{A}\}\odot\{\text{B}\}$ &  $\{\text{A}\}+\{\text{B}\}$   &     $\{\text{A}\}-\{\text{B}\}$      \\ \hline
			0     &     0     &                0                &               0                &                  0                   \\
			0     &     1     &                0                &               1                &                $(-1)$                \\
			1     &     0     &                0                &               1                &                  1                   \\
			1     &     1     &                1                &              (2)               &                  0
		\end{tabular}
	\end{ruledtabular}
\end{table}


\section{Discussion and outline}

There are multiple possible ways to extend the presented formalism. The mapping will work similar in space with more antisymmetries~\cite{Padmanabhan2020}, like rotation-reversal~\cite{Gopalan2011, VanLeeuwen2014, Huang2014}, distortion reversal~\cite{VanLeeuwen2015}, wedge reversal~\cite{Gopalan2020, Fabrykiewicz2023b} or charge conjugation~\cite{Hitzer2024}. With more antisymmetries there will be more types of physical properties~\cite{Gopalan2020, Fabrykiewicz2023b}, but the mapping still will be an epimorphism with a kernel consisting of all antisymmetries to which the selected property with given type is invariant. It is essential that for any number of antisymmetries there will be only 32 PRPGs. The mapping will also work similar in space with different dimensions than three~\cite{Shubnikov_Koptsik_book, Souvignier2006, Gopalan2020, Fabrykiewicz2023b}. 

Bidirectors~\cite{Hlinka2014, Erb2020} have length and orientation like vectors, but none of their ends is distinguished. The minimal non-zero angle rotation symmetry about axis perpendicular to distinguished direction for spinors, vectors and bidirectors is $4\pi$, $2\pi$ and $\pi$, respectively. Bidirectors are denoted as \textbf{N}, \textbf{L}, \textbf{C}, \textbf{F} for even-, magnetic-, electric- and magnetoelectric-types, respectively~\cite{Hlinka2014, Erb2020}. As discussed in~\cite{Hlinka2014} bidirectors correspond to the symmetric tensors of rank~2 with only a single nonzero element in its canonical form. By expressing tensor of $xG^2$ intrinsic symmetry in terms of multipoles (see Appendix), one will find that such form correspond to $X_0=X_u$ with zero other multipole components~\footnote{Both $1^\dagger$-odd bidirector and $X_u$ component of $1^\dagger$-odd quadrupole have $\infty2$ symmetry, which is a subgroup of $X_0$ $1^\dagger$-odd monopole $\infty\infty$ symmetry (see Table S4 in SM~\cite{SM_joint}), so always $X_0$ is allowed when $X_u$ is allowed.}, related to the sd orbital hybridization~\cite{Landis1995, Landis1996, Landis1998, Landis1999, Firman2001, Landis2006, Fekl2021, Borge-Duran2021, Mao2025}. Bidirectors were discussed in the context of macroscopic properties~\cite{Hlinka2014} and their correspondence to multipoles and sd orbital hybridization opens a possibility to use them for description of microscopic atomic-scale properties.

The main advantage of the proposed method is that the incompletely investigated relations and operations drawn with dashed arrows in Fig.~\ref{Fig_graph} are now fully explained. With this result the relations between groups can be extrapolated to relations between symmetry allowed physical properties.

\begin{acknowledgments}
	Thanks are due to Martin Meven, Leonie St\"odter, Micha\l{} St\k{e}kiel (Forschungszentrum J\"ulich GmbH), Rados\l{}aw Przenios\l{}o and Izabela Sosnowska (University of Warsaw) for inspiring discussions.
\end{acknowledgments}

\textit{Data availability}$-$ All data are included within the Letter and its Supplemental Material.

\section{Appendix}

This appendix gives examples of $xG^2$ and $xG[G^2]$ tensors expressed in terms of multipole degrees of freedom.

The scalar ($x$ tensor) and vector ($xG$ tensor) are equivalent to the monopole, $(X_0)$, and the dipole $(X_x, X_y, X_z)$, respectively. The $xG^2$ tensor corresponds to the monopole, $(X_0)$, dipole, $(X_x, X_y, X_z)$ and quadrupole, $(X_u, X_v, X_{yz}, X_{zx}, X_{xy})$:
\begin{equation}
	\chi_{xG^2} =
	\begin{bmatrix}
		X'_0 + X_v   & X_{xy} + X_z & X_{zx} - X_y \\
		X_{xy} - X_z & X'_0 - X_v   & X_{yz} + X_x \\
		X_{zx} + X_y & X_{yz} - X_x & X'_0 +  X'_u
	\end{bmatrix}
\end{equation}
where $X'_0 = X_0 - X_u$ and $X'_u = 3 X_u$. For determining tensor form, one can treat $X'_0$ as a monopole, because it is impossible that $X_u$ is allowed but $X_0$ is symmetry prohibited, as $X_u$ group ($\infty_z 2$) is a subgroup of $X_0$ group ($\infty\infty$). Symmetric $x[G^2]$ and antysymmetric $x\{G^2\}$ parts of the $xG^2$ tensor corresponds to monopole, $(X_0)$, plus quadrupole, $(X_u, X_v, X_{yz}, X_{zx}, X_{xy})$ and alone dipole, $(X_x, X_y, X_z)$, respectively.

The $xG[G^2]$ tensor expressed in cubic harmonics corresponds to two independent dipoles: $(X_x, X_y, X_z)$ and $(Y_x, Y_y, Y_z)$, quadrupole, $(X_u, X_v, X_{yz}, X_{zx}, X_{xy})$ and octupole: $(X_{xyz}, X_z^\alpha, X_z^\beta, X_x^\alpha, X_x^\beta, X_y^\alpha, X_y^\beta)$:
\begin{equation}
	\label{Eq_xG[G2]}
	\chi_{xG[G^2]} =
	\begin{bmatrix}
		X'_x           & X''_y - A_y    & X''_z + A_z     \\
		X''_x + A_x    & X'_y           & X''_z - A_z     \\
		X''_x - A_x    & X''_y + A_y    & X'_z            \\
		X'_{xyz} + X_u & X'''_z - B_z   & X'''_y + B_y    \\
		X'''_z + B_z   & X'_{xyz} - X_u & X'''_x - B_x    \\
		X'''_y - B_y   & X'''_x + B_x   & X'_{xyz} + X'_v
	\end{bmatrix}^T,
\end{equation}
where $X'_{xyz} = X_{xyz} + X_v$, $X'_v = -3 X_v$, $(X'_x, X''_x, X'''_x) = (Y_x - 4 X_x + 4 X_x^\alpha, Y_x + 2 X_x - 2 X_x^\alpha, - 3 X_x - 2 X_x^\alpha)$ (cyclic) and $(A_x, B_x) = (2X_x^\beta + 2X_{yz}, 2X_z^\beta - X_{yz})$ (cyclic). For determining tensor form, one can treat $X'_{xyz}$ as a $X_{xyz}$ quadrupole component, because it is impossible that $X_v$ is allowed but $X_{xyz}$ is symmetry prohibited, as $X_v$ group ($4^\dagger22^\dagger$) is a subgroup of $X_{xyz}$ group ($4^\dagger32^\dagger$). For determining tensor form, one can treat $X'_i$, $X''_i$ and $X'''_i$ with $i=x,y,z$ as a three independent $X_i$ dipole components, because $Y_i$, $X_i$ and $X_i^\alpha$ have the same $\infty_i2^\dagger$ symmetry. The form of $x[G^2]G$ tensors is equal to transposed form of $xG[G^2]$. The $xG[G^2]$ tensor in tesseral harmonics as well as $x[G^3]$, $xG[G^3]$, $x[G^2][G^2]$ tensors expressed in both cubic and tesseral harmonics can be found in~\cite{Yatsushiro2021, Yatsushiro2022}.


\bibliography{Tensors.bib}

\end{document}